# Phonon transport in single-layer Boron nanoribbons

Zhongwei Zhang[1], Yuee Xie[1], Qing Peng[2], and Yuanping Chen[1*]

[1] Department of Physics, Xiangtan University, Xiangtan 411105, Hunan, P.R. China

[2] Department of Mechanical, Aerospace and Nuclear Engineering, Rensselaer Polytechnic Institute, Troy, NY, 12180, USA

*Corresponding author: chenyp@xtu.edu.cn

**Abstract:** Inspired by the successful synthesis of several allotropes, boron sheets have been one of the hottest spot areas of focus in various fields. Here, we study phonon transport in three types of boron nanoribbons with zigzag and armchair edges by using a non-equilibrium Green's function combined with first principles methods. Diverse transport properties are found in the nanoribbons. At the room temperature, their highest thermal conductance can be comparable with that of graphene, while the lowest thermal conductance is less than half of graphene's. The three boron sheets exhibit different anisotropic transport characteristics. Two of these sheets have stronger phonon transport abilities along the zigzag edges than the armchair edges, while in the case of the third, the results are reversed. With the analysis of phonon dispersion, bonding charge density, and simplified models of atomic chains, the mechanisms of the diverse phonon properties are discussed. Because all boron allotropes consists of hexagonal and triangular rings, many hybrid patterns can be constructed naturally without doping, adsorption, and defects. Our results are useful in materials and devices design using boron sheets for thermal management.

Thermal management is critical in electronics. Graphene is considered to be an outstanding thermal material because of its super high thermal conductivity.[1-4] In thermal devices based on graphene, phonon transport needs to be modulated to obtain different functions and applications, such as phonon rectifiers,[5,6] phonon filters,[7] phonon transistors[8,9], and thermoelectric devices.[10,11] Heterostructures and periodic patterns are frequently used modulation structures, which can be obtained by adsorption,[12,13] doping,[14] and defects.[15,16] For example, periodic linear or rectangular patterns can be created by regular hydrogen adsorption;[10,12] antidot periodic patterns can be formed by vacancy defects.[16,17] It has been proven that the periodic patterns indeed tune phonon transport effectively.[11-13,16] However, to obtain the different patterns, the lattice of graphene must be destroyed and accurate experimental conditions are required.[18,19]

Boron is another element possessing rich chemical properties in addition to carbon.[20-22] Many two-dimensional (2D) single layer boron allotropes have been predicted theoretically, such as $\alpha$-, $\beta$-, $g$-boron.[21-23] In these boron sheets, boron exhibits more diverse bonds than carbon. It is known that, in the 2D carbon sheets, one carbon atom may have 2, 3, and 4 bonds with its nearest neighboring atoms, while one boron atom can have 3 to 7 bonds connected with other boron atoms.[21,24,25] Because boron atoms appear to form triangular or hexagonal rings, almost all boron allotropes have close lattice geometries, which can be considered as inserting atoms to parts of hexagonal rings in a honeycomb lattice or removing parts of atoms from a triangular lattice.[21,26,27] A parameter, the so-called hexagonal vacancy density $\eta$,[21,23] is used to describe the ratio of hexagon holes to the number of atomic sites in the original triangular lattice within one unit cell. Boron allotropes with different $\eta$ show different electronic properties, such as high anisotropy[21,28] and superconductivity,[29,30] which have attracted much attention. More inspiringly, several boron allotropes have been synthesized successfully.[28,31] Feng *et al.* obtained $\beta_{12}$ and $\chi_3$ using molecular beam epitaxy on a Ag (111) surface.[31] A triangular boron sheet was reported by Andrew *et al.* in their experiment.[28] This further suggests boron materials to be one of the hottest areas of research in various fields. However as far as we know, phonon transport in boron sheets has not been reported to date. An intriguing question is: what is the phonon transport ability of these 2D boron sheets? Moreover, boron allotropes consisting of triangular or hexagonal rings can form various patterns without destroying their pristine lattices, and thus provide rich hybrid structures to modulate thermal transport.

In this work, we study phonon transport in boron nanoribbons curt from three types of boron sheets, as shown in Fig. 1, by using a non-equilibrium Green's function (NEGF) combined with first principles methods.[32-35] The three boron sheets show different phonon transports along zigzag and armchair directions, and the transport mechanisms are explained by phonon dispersion, bonding charge densities, and simplified models of atomic chains. Based on the three boron sheets, various hybrid boron structures with different patterns are proposed, and their potential applications are also suggested.

## Model and Simulation methods

Three types of boron sheets are considered, as show in Figs. 1(b)-1(d), named boron 1, 2, and 3 respectively. Boron 1and 2 have been synthesized successfully,[31] while boron 3 is a boron allotrope that is highly stable even if it has not been reported to have been experimentally synthesized.[23,36] All of them can be viewed as boron structures by periodically inserting atoms to honeycomb or periodically removing atoms from triangular lattices. Different insertion or removal methods lead to distinct patterns and atomic densities of boron allotropes. To study the phonon transport properties along different edges and directions, 2D sheets are always cut into quasi-one-dimensional nanoribbons. For example, graphene has two typical nanorbbons: zigzag-edged nanoribbons (G-ZNR) and armchair-edged nanoribbons (G-ANR). Similarly, boron 1/2/3 can also be cut into zigzag- and armchair-edged nanoribbons, named as B1/2/3-ZNR and B1/2/3-ANR, respectively. The two types of boron nanoribbons are denoted by solid atoms in Figs. 1(b-d), whose widths are labeled as $W_Z$ and $W_A$, respectively.

Phonon transport properties of the boron nanoribbons can be calculated by a NEGF method combined with first principles calculations. Each nanoribbon can be divided into three parts: left lead, right lead, and the center scattering region. According to the NEGF scheme, the retarded Green's function of the nanoribbon is expressed as[33,34,37-39]

$$G^r = \left[(\omega + i0^+)^2 I - K^C - \sum\nolimits_L^r - \sum\nolimits_R^r\right]^{-1}, \tag{1}$$

where $\omega$ is the frequency of phonons, $K^C$ is the mass-weighted force constants matrix of the center region, and $\sum_\beta^r = V^{C\beta}\, g_\beta^r V^{\beta C}$ ($\beta = L, R$, corresponding to the left and right) denotes the self-energy of the (left or right) lead $\beta$, in which $V^{C\beta} = (V^{\beta C})^T$ is the coupling matrix of the lead $\beta$ to the center region and $g_\beta^r$ is the surface Green's function of the lead. Once the retarded Green's function $G^r$ is obtained, we can calculate transmission coefficient $T[\omega]$ and then the thermal conductance $\kappa$ of the nanoribbon:

$$T[\omega] = Tr\{G^r \Gamma_L G^a \Gamma_R\}, \tag{2}$$

$$\kappa(T) = \frac{\hbar}{2\pi}\int_0^\infty T[\omega]\omega \frac{\partial f(\omega)}{\partial T} d\omega, \tag{3}$$

where $\Gamma_\beta = i\left(\sum_\beta^r - \sum_\beta^a\right) = -2ImV^{\beta C} g_\beta^r V^{\beta C}$ is the coupling function of the $\beta$ lead, and $f(\omega) = \left\{exp\left[\frac{\hbar\omega}{\kappa_B T}\right] - 1\right\}^{-1}$ is the Bose-Einstein distribution function for heat carriers at the leads. In addition, the phonon spectrum of the nanoribbons can be obtained from the generalized eigenvalue method:

$$\begin{pmatrix} \omega^2 I - K_{11} & I \\ K_{10} & 0 \end{pmatrix}\begin{pmatrix} \varepsilon \\ \xi \end{pmatrix} = \lambda \begin{pmatrix} K_{01} & 0 \\ 0 & I \end{pmatrix}\begin{pmatrix} \varepsilon \\ \xi \end{pmatrix}, \tag{4}$$

After diagonalizing this generalized eigenvalue matrix, one can get the eigenvalues $\lambda$. By setting the traveling waves eigenvalue to be ($\lambda = e^{iqa}$), the wave number $q$ for a special $\omega$ is found. And then, the phonon spectrum of the nanoribbons is obtained.

The force constant $K^C$ in Eq. (1) can be calculated from first principles method based on the following form:[40,41]

$$K_{xy}^{ij} = \frac{\partial^2 u_{ij}}{\partial x \partial y}, \tag{5}$$

where $K_{xy}^{ij}$ is the force constant between the *i*th and *j*th neighboring atoms in the *x* and *y* directions, and $u_{ij}$ is the potential energy between the *i*th and *j*th atoms, respectively. That is, the force constant $K_{xy}^{ij}$ is calculated from the second derivative of potential energy with small displacements in $\partial x$ and $\partial y$. In order to accurately describe the interactions between atoms, the considered force constants are up to the third neighboring atom in each direction.

## Results and Discussions

In Figs. 2(a) and 2(b), the thermal conductance $\kappa$ for ZNRs and ANRs of the three types of boron sheets as a function of temperature *T* are shown ($W_{Z/A} \approx 5.0$ nm), respectively. The $\kappa$ for G-ZNR and G-ANR are also shown for comparison. All the $\kappa$ curves increase with $T$ because the phonon modes are gradually excited by high temperatures. Both G-ANR and G-ZNR have higher $\kappa$ values than those of boron nanoribbons. The three types of boron sheets exhibit diverse phonon transport abilities. For the ZNRs, the $\kappa$ of B1-ZNR is the highest, while that of B3-ZNR is the lowest, as shown in Fig. 2(a). Moreover, the former is nearly two times of the latter. For the ANRs, the $\kappa$ values of the three boron sheets have little difference (see Fig. 2(b)). This implies that the three boron sheets possess a completely different transport anisotropy.

Previous studies reported that 2D materials have different dependence relationships between $\kappa$ and $T$.[37,42-44] For example, many 2D materials exhibit $T$ and $T^2$ dependences of $\kappa$,[42,43] while a $T^{1.5}$ dependence has been found in graphene and BN.[37,42,44] We fit the $\kappa$ curves of boron nanoribbons in Figs. 2(a) and 2(b) by the following formula:

$$\kappa(W,T) = (a_1 + b_1 W)T + (a_2 + b_2 W)T^{1.5} + (a_3 + b_3 W)T^2, \qquad (6)$$

where $a_i$ and $b_i$ are parameters. The optimal fitting parameters for the boron nanoribbons are given in Table 1. It is found that, similar to graphene, the $T^{1.5}$ term in Eq. (6) has an important contribution to $\kappa$. The detailed analysis of the contributions of $T$, $T^{1.5}$, and $T^2$ is given in the Supplementary Information (SI).

Figures 2(c) and 2(d) present the dependences of $\kappa$ on the width $W_{\mathrm{Z/A}}$ of boron ZNRs and ANRs ($T$ = 300K), respectively. Because the number of phonon channels depends linearly on the widths of the nanoribbons, all the $\kappa$ values vary linearly as $W_{\mathrm{Z/A}}$ increases. However, the slopes of the lines are different. B1-ZNR has the largest slope that is close to graphene, indicating that its $\kappa$ values are close to graphene's at the room temperature. The slope of the B3-ZNR is the smallest, and thus the difference in $\kappa$ between B3-ZNR and B1-ZNR increases with the width. However, the difference in $\kappa$ between boron ANRs is very small with an increase in width. Therefore, the anisotropies of $\kappa$ in the boron sheets will increase with the widths.

To directly compare the anisotropies of intrinsic transport properties of boron nanoribbons, the scaled thermal conductivities ($\sigma/L$) for both ZNRs and ANRs at 300K are shown in Fig. 3. Thermal conductivity $\sigma$ is calculated by $\sigma = \kappa/S \times L$, where $S = W_{Z/A} \times h$ is the cross sectional area ($h$ = 0.32 nm is the thickness of boron sheets)[45] and $L$ is the mean free path of phonons. Because $L$ has not been reported, here we calculate the $L$ scaled thermal conductivity. One can find from Fig. 3 that, with an increase in width, the thermal conductivities drop at first and then approach converged values. The different transport abilities of the ZNRs are shown again (see Fig. 3(a)), while the difference between the ANRs is small (see Fig. 3(b)). We define $\varepsilon = 1 - (\sigma/L)_{ANR}/(\sigma/L)_{ZNR}$ to indentify the anisotropic degree of phonon transport in different boron sheets. Then, $\varepsilon$ =39.1%, 47.2%, and -14.4% for boron 1, 2, and 3, respectively. This indicates that boron 1 and 2 possess very strong transport anisotropies ($\varepsilon \approx 20\%$ for graphene).[46] The thermal conductivities of the ZNRs are much higher than those of the ANRs. Interestingly, the anisotropy of boron 3 is a negative value, i.e., thermal conductivities of ANRs are higher than those of ZNRs. The different transport characteristics on one hand demonstrate wide ranges of thermal conductivities of boron sheets, on the other hand show diverse transport phenomena.

Phonon transport properties are related to the phonon dispersions. To explain the difference of phonon transport in the boron sheets, in Fig. 4, the phonon spectra of the boron nanoribbons and also graphene nanoribbons are given. The comparison of the phonon spectra reveals that the frequency ranges and the acoustic phonon velocities determine transport abilities. For example, the phonon dispersions of B1-ZNR are much closer to those of G-ZNR, and thus it has the highest thermal conductivity in all boron nanoribbons. While the lower thermal conductivity of B3-ZNR is because of its small group velocities of both optical and acoustic phonons. The different anisotropies of boron sheets are also reflected by the spectra. The group velocities of acoustic phonons in the B1- and B2-ZNRs are obviously faster than those in the ANRs, while the case of boron 3 is opposite. Therefore, the anisotropies of the former two are opposite to that of the third one. In addition, it is noted that there exist quadratic acoustic branches in the spectra, which are responsible for the $T^{1.5}$ dependence in Eq. (6).[42]

The different phonon transport properties in the boron sheets can be understood from their bonding charge densities (BCD),[47] which are the charge density difference between the valence charge density of the system and the superposition of the valence charge density of the neutral constituent atoms. Figures 5(a-d) show BCD contours of ground state graphene and boron 1, 2, and 3, respectively. They directly reflect the strength of interactions between atoms. In Fig. 5(a), electrons are evenly distributed between the carbon atoms and thus form a honeycomb geometry, which has been explained as an intrinsic reason of the high conductivity of graphene.[42,44] The BCD contour of boron 1 is somewhat similar to that of graphene except the interactions are weaker, as shown in Fig. 5(b). Therefore, its phonon transport properties are mostly close to those of graphene. Seen from Figs. 5(c) and 5(d), boron 2 and 3 have completely different contours. In boron 2, the BCD contour exhibits strong directionality: along the zigzag direction the contour forms channels while the channels are truncated along the armchair direction. This is the cause for the high transport anisotropy. In boron 3, the contour forms channels along both zigzag and armchair directions. However, the channels along the armchair direction are linear, while those along zigzag direction are zigzag. The former are more advantageous to transport, and thus the thermal conductivity along the armchair direction is higher than that along the zigzag direction. Consequently, boron 3 has inverse anisotropy compared to boron 1 and 2.Meanwhile, the contour of boron 3 implies that the interactions between atoms are weaker, which results in low acoustic phonon velocities and thermal conductivity.

To further understand the relations between pattern geometries and phonon transport properties, we simplify the boron nanoribbons to single atomic chains or two coupled atomic chains. For comparison, the simplified models of graphene nanoribbons are first given in the inset of Fig. 6. Both G-ZNR and G-ANR can be simplified to monatomic chains, because all hexagonal rings are equivalent. However, the hexagonal rings in the boron nanoribbons are not equivalent any more, and thus complicated atomic chains are needed to describe their properties. For example, in born 1, its ZNR is simplified to a monatomic chain whereas AGR is simplified to a diatomic chain, because the hexagonal rings with and without inserting atoms alternately appear along the armchair direction. The case of boron 2 is the same to boron 1. As to the boron 3, both ZNR and ANR are simplified to two coupled diatomic chains because of their relatively complicated periodic geometries. On the basis of atomic chains of graphene, the atomic masses $m$ and lattice force constants $K$ in the boron atomic chains are figured out (see the insets in Fig. 6). Figure 6 shows $\kappa$ values of the corresponding simplified models. One can find that the simplified models reproduce the phonon transport characteristics of the boron nanoribbons. This not only explains the origination of diverse transport properties in the three types of boron sheets, but also provides an effective method to predict phonon properties of boron allotropes with other patterns.

As mentioned above, with the character that boron allotropes consisting of triangular or hexagonal rings can form various patterns without destroying their pristine lattices. In Fig. 7, some patterns based on boron 1, 2 and 3 are proposed and several potential applications are suggested. Figures 7(a-c) present perpendicular, parallel and array patterns constructed by boron 1 and 3, and Fig. 7(d) presents a hybrid pattern constructed by three types of boron nanoribbons. Because the transport properties of boron 1, 2 and 3 are different, these patterns will modulate thermal transport in different ways, and thus various functions and related devices are expected to be obtained. In Fig. 7(e), a triangular heterostructure is constructed by boron 1 and 3.Itcan be served as a phonon rectifier, because the big asymmetry of geometry and the difference of phonon properties between two sides.  Figure 7(f) shows a three-terminal T-shaped junction consisting of boron 1, 2 and 3. It may be a potential phonon transistor where the boron 1 and 2 are two leads while the boron 3 is served as a gate. Besides boron 1, 2 and 3, there are many other boron allotropes consisting of triangular or hexagonal rings.[21,23,24] Therefore, many hybrid patterns can be constructed naturally without doping, adsorption and defects, to fulfill the demands of thermal management.

In summary, we have studied phonon transport in three types of boron sheets with different edges, using a combined method of NEGF and first principles calculations. The boron sheets exhibit diverse phonon transport abilities and different anisotropies. The highest thermal conductivity is comparable to that of graphene, while the lowest thermal conductivity is less than half of graphene's. Moreover, boron 1 and 2 exhibit stronger transport abilities along the zigzag edges than the armchair edges, while the case of boron 3 is opposite. The phonon dispersions and BCD contours reveal the originations of the transport properties. The boron nanoribbons are simplified to atomic chains to further explore the difference between the transport properties, which provides an effective method to predict phonon transport of other boron allotropes. Hybrid patterns based on the boron sheets are constructed. All the hybrid structures have natural interfaces without doping, adsorption and defects. Some potential applications of hybrid boron sheets in thermal devices, such as rectifier and transistors, are proposed. With the analytic formula of the thermal conductivity with respect to the width and temperature of these three base units, the thermal conductance of the hybrid structures are ready to predict. Our study is useful in material and devices design for thermal management.

**Acknowledgments**

This work was supported by the National Natural Science Foundation of China (Nos. 51376005 and 11474243).

**Notes**

The authors declare no competing financial interest.

**References**

(1) Balandin, A. A.; Ghosh, S.; Bao, W.; Calizo, I.; Teweldebrhan, D.; Miao, F.; Lau, C. N., *Nano Lett* **2008,** *8*, 902-907.

(2) Ghosh, S.; Calizo, I.; Teweldebrhan, D.; Pokatilov, E. P.; Nika, D. L.; Balandin, A. A.; Bao, W.; Miao, F.; Lau, C. N., *Appl Phys Lett* **2008,** *92*, 151911.

(3) Balandin, A. A., *Nat Mater* **2011,** *10*, 569-581.

(4) Huang, X.; Yin, Z.; Wu, S.; Qi, X.; He, Q.; Zhang, Q.; Yan, Q.; Boey, F.; Zhang, H., *Small* **2011,** *7*, 1876-1902.

(5) Hu, J.; Ruan, X.; Chen, Y. P., *Nano Lett* **2009,** *9*, 2730-2735.

(6) Ouyang, T.; Chen, Y.; Xie, Y.; Wei, X. L.; Yang, K.; Yang, P.; Zhong, J., *Phys Rev B* **2010,** *82*, 245403.

(7) Kaike, Y.; Yuanping, C.; Yuee, X.; Tao, O.; Jianxin, Z., *Euro Phys Lett* **2010,** *91*, 46006.

(8) Li, B.; Wang, L.; Casati, G., *Appl Phys Lett* **2006,** *88*, 143501.

(9) Pop, E.; Varshney, V.; Roy, A. K., *MRS Bulletin* **2012,** *37*, 1273-1281.

(10) Kim, J. Y.; Grossman, J. C., *Nano Lett* **2015,** *15*, 2830-2835.

(11) Yang, K.; Chen, Y.; D'Agosta, R.; Xie, Y.; Zhong, J.; Rubio, A., *Phys Rev B* **2012,** *86*, 045425.

(12) Pei, Q.-X.; Sha, Z.-D.; Zhang, Y.-W., *Carbon* **2011,** *49*, 4752-4759.

(13) Zhang, Z.; Xie, Y.; Peng, Q.; Chen, Y., *Solid State Commun* **2015,** *213-214*, 31-36.

(14) Chen, S.; Wu, Q.; Mishra, C.; Kang, J.; Zhang, H.; Cho, K.; Cai, W.; Balandin, A. A.; Ruoff, R. S., *Nat Mater* **2012,** *11*, 203-207.

(15) Feng, T.; Ruan, X.; Ye, Z.; Cao, B., *Phys Rev B* **2015,** *91*, 224301.

(16) Ng, T. Y.; Yeo, J. J.; Liu, Z. S., *Carbon* **2012,** *50*, 4887-4893.

(17) Fthenakis, Z. G.; Zhu, Z.; Tománek, D., *Phys Rev B* **2014,** *89*, 125421.

(18) Liu, Z.; Ma, L.; Shi, G.; Zhou, W.; Gong, Y.; Lei, S.; Yang, X.; Zhang, J.; Yu, J.; Hackenberg, K. P.; Babakhani, A.; Idrobo, J. C.; Vajtai, R.; Lou, J.; Ajayan, P. M., *Nat Nanotechnol* **2013,** *8*, 119-124.

(19) Sessi, P.; Guest, J. R.; Bode, M.; Guisinger, N. P., *Nano Lett* **2009,** *9*, 4343-4347.

(20) Albert, B.; Hillebrecht, H., *Angew Chem Int Ed* **2009,** *48*, 8640-8668.

(21) Wu, X.; Dai, J.; Zhao, Y.; Zhuo, Z.; Yang, J.; Zeng, X. C., *ACS Nano* **2012,** *6*, 7443-7453.

(22) Tang, H.; Ismail-Beigi, S., *Phys Rev Lett* **2007,** *99*, 115501.

(23) Tang, H.; Ismail-Beigi, S., *Phys Rev B* **2010,** *82*, 115412.

(24) Liu, Y.; Penev, E. S.; Yakobson, B. I., *Angew Chem Int Ed Engl* **2013,** *52*, 3156-3159.

(25) Özdoğan, C.; Mukhopadhyay, S.; Hayami, W.; Güvenç, Z. B.; Pandey, R.; Boustani, I., *J Chem Phys C* **2010,** *114*, 4362-4375.

(26) Liu, H.; Gao, J.; Zhao, J., *Sci Rep* **2013,** *3*, 3238.

(27) Lau, K. C.; Pandey, R., *J Phys Chem B* **2008,** *112*, 10217-10220.

(28) Mannix, A. J.; Zhou, X.-F.; Kiraly, B.; Wood, J. D.; Alducin, D.; Myers, B. D.; Liu, X.; Fisher, B. L.; Santiago, U.; Guest, J. R.; Yacaman, M. J.; Ponce, A.; Oganov, A. R.; Hersam, M. C.; Guisinger, N. P., *Science* **2015,** *350*, 1513-1516.

(29) Penev, E. S.; Kutana, A.; Yakobson, B. I., *Nano Lett* **2016,** *16*, 2522-2526.

(30) Zhao, Y.; Zeng, S.; Ni, J., *Phys Rev B* **2016,** *93*, 014502.

(31) Feng, B.; Zhang, J.; Zhong, Q.; Li, W.; Li, S.; Li, H.; Cheng, P.; Meng, S.; Chen, L.; Wu, K., *Nat Chem* **2016**, doi:10.1038/nchem.2491.

(32) Blase, X.; Charlier, J.-C.; De Vita, A.; Car, R., *Appl Phys Lett* **1997,** *70*, 197-199.

(33) Mingo, N., *Phys Rev B* **2006,** *74*, 125402.

(34) Yamamoto, T.; Watanabe, K., *Phys Rev Lett* **2006,** *96*, 255503.

(35) Ouyang, T.; Hu, M., *Nanotechnol* **2014,** *25*, 245401.

(36) Peng, Q.; Han, L.; Wen, X.; Liu, S.; Chen, Z.; Lian, J.; De, S., *Phys Chem Chem Phys* **2015,** *17*, 2160-2168.

(37) Ouyang, T.; Chen, Y.; Xie, Y.; Yang, K.; Bao, Z.; Zhong, J., *Nanotechnol* **2010,** *21*, 245701.

(38) Ouyang, T.; Chen, Y.; Xie, Y.; Stocks, G. M.; Zhong, J., *Appl Phys Lett* **2011,** *99*, 233101.

(39) Zhang, Z.; Xie, Y.; Peng, Q.; Chen, Y., *Sci Rep* **2016,** *6*, 21639.

(40) Kresse, G.; Furthmüller, J., *Comput Mater Sci* **1996,** *6*, 15-50.

(41) Togo, A.; Oba, F.; Tanaka, I., *Phys Rev B* **2008,** *78*, 134106.

(42) Mingo, N.; Broido, D. A., *Phys Rev Lett* **2005,** *95*, 096105.

(43) Hong, Y.; Zhang, J.; Huang, X.; Zeng, X. C., *Nanoscale* **2015,** *7*, 18716-18724.

(44) Jiang, J.-W.; Wang, J.-S.; Li, B., *Phys Rev B* **2009,** *79*, 205418.

(45) Liu, F.; Shen, C.; Su, Z.; Ding, X.; Deng, S.; Chen, J.; Xu, N.; Gao, H., *J Mater Chem* **2010,** *20*, 2197-2205.

(46) Xu, Y.; Chen, X.; Gu, B.-L.; Duan, W., *Appl Phys Lett* **2009,** *95*, 233116.

(47) Ouyang, T.; Hu, M., *Phys Rev B* **2015,** *92*, 235204.

## Figure Captions:

**Figure 1.** Atomic structures of graphene and three types of boron sheets: graphene (a), boron 1 (b), boron 2 (c), and boron 3 (d). The solid atoms between the red and blue lines represent ZNRs and ANRs, respectively, whose widths are labeled as $W_Z$ and $W_A$. The dark green atoms in (b), (c), and (d) represent the inserting atoms in the honeycomb structure.

**Figure 2.** Thermal conductance$\kappa$ of (a) G-ZNR, B1-ZNR, B2-ZNR, and B3-ZNR, and (b) G-ANR, B1-ANR, B2-ANR, and B3-ANR as a function of temperature at $W_{Z/A} \approx 5.0$ nm, respectively. Thermal conductance $\kappa$ of (c) G-ZNR, B1-ZNR, B2-ZNR, and B3-ZNR, and (d) G-ANR, B1-ANR, B2-ANR, and B3-ANR as a function of widths $W_{Z/A}$ at $T$ = 300K, respectively.

**Figure 3.** (a) Scaled thermal conductivities $\sigma/L$ of boron ZNRs as a function of $W_Z$ at $T$ = 300K. (b) Scaled thermal conductivities $\sigma/L$ of boron ANRs as a function of $W_A$ at $T$ = 300K. The inserted blank dots represent the corresponding values of graphene nanoribbons.

**Figure 4.** Phonon spectrums of (a) G-ZNR and G-ANR, (b) B1-ZNRand B1-ANR, (c) B2-ZNR and B2-ANR, and (d) B3-ZNR and B3-ANR, respectively.

**Figure 5.** BCD contours for (a) graphene, (b) boron 1, (c) boron 2, and (d) boron 3, respectively. Black circles represent the positions of atoms and black lines represent strong bonds.

**Figure 6.** Normalized $\kappa$ of simplified models of atomic chains corresponding to graphene and boron nanoribbons. The inset figures are simplified models of atomic chains where the parameters of mass $m$ and force constant $K$ are shown.

**Figure 7.** Atomic structures of hybrid boron patterns. (a) Perpendicular pattern, (b) parallel pattern, and (c) array pattern consisting of boron 1 and 3. (d) Multi-Perpendicular pattern consisting of boron 1, 2, and 3. (e) A phonon rectifier with triangular shape based on boron 1 and 3. (f) A three-terminal phonon transistor based on boron 1, 2, and 3.

**Table 1.** The values of fitting parameters $a_i$ and $b_i$ (i =1, 2, 3) in Eq. (6).

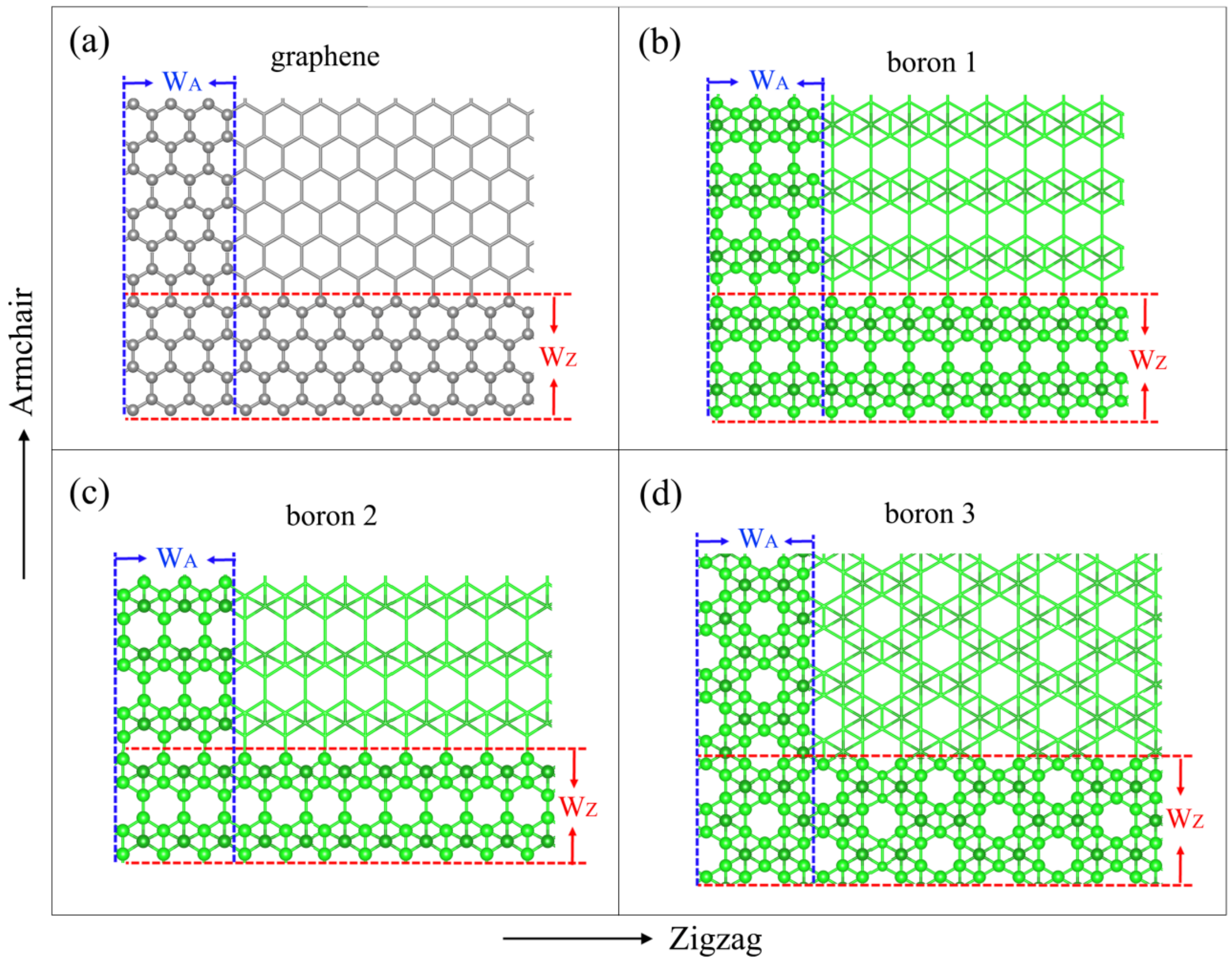

**Figure 1.** Atomic structures of graphene and three types of boron sheets: graphene (a), boron 1 (b), boron 2 (c), and boron 3 (d). The solid atoms between the red and blue lines represent ZNRs and ANRs, respectively, whose widths are labeled as $W_Z$ and $W_A$. The dark green atoms in (b), (c), and (d) represent the inserting atoms in the honeycomb structure.

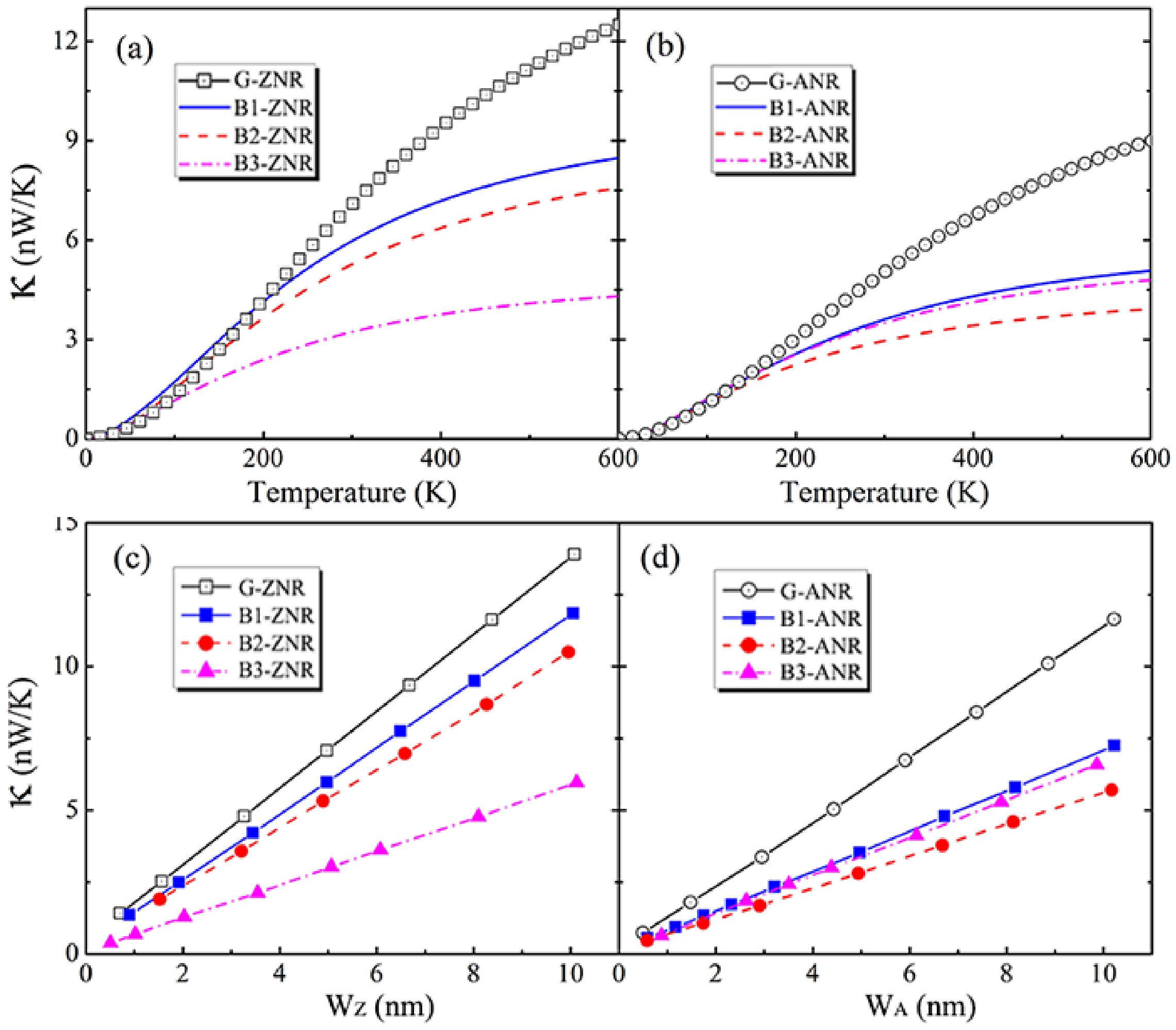

**Figure 2.** Thermal conductance $\kappa$ of (a) G-ZNR, B1-ZNR, B2-ZNR, and B3-ZNR, and (b) G-ANR, B1-ANR, B2-ANR, and B3-ANR as a function of temperature at $W_{Z/A} \approx 5.0$ nm, respectively. Thermal conductance $\kappa$ of (c) G-ZNR, B1-ZNR, B2-ZNR, and B3-ZNR, and (d) G-ANR, B1-ANR, B2-ANR, and B3-ANR as a function of widths $W_{Z/A}$ at $T$ = 300K, respectively.

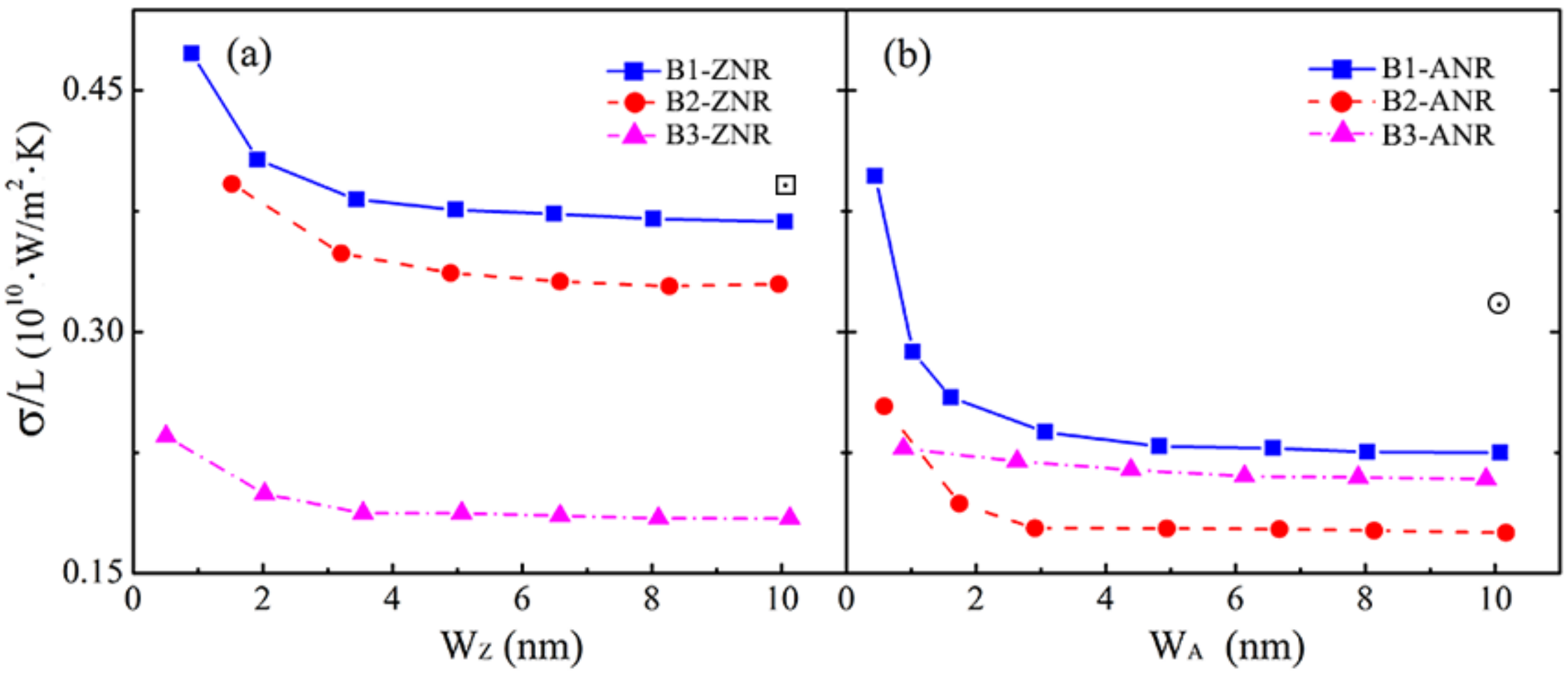

**Figure 3.** (a) Scaled thermal conductivities $\sigma/L$ of boron ZNRs as a function of $W_Z$ at $T$ = 300K. (b) Scaled thermal conductivities $\sigma/L$ of boron ANRs as a function of $W_A$ at $T$ = 300K. The inserted blank dots represent the corresponding values of graphene nanoribbons.

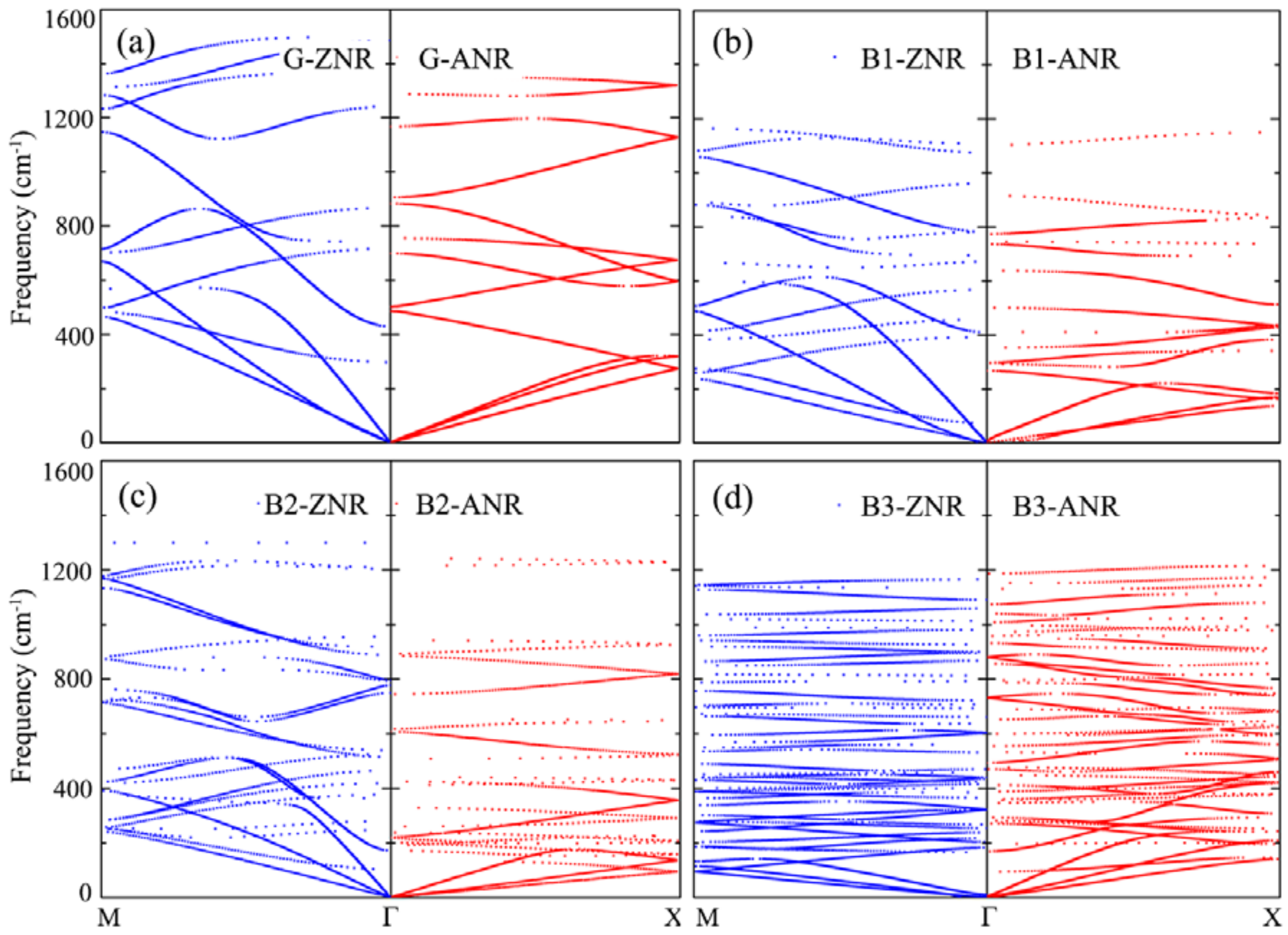

**Figure 4.** Phonon spectrums of (a) G-ZNR and G-ANR, (b) B1-ZNRand B1-ANR, (c) B2-ZNR and B2-ANR, and (d) B3-ZNR and B3-ANR, respectively.

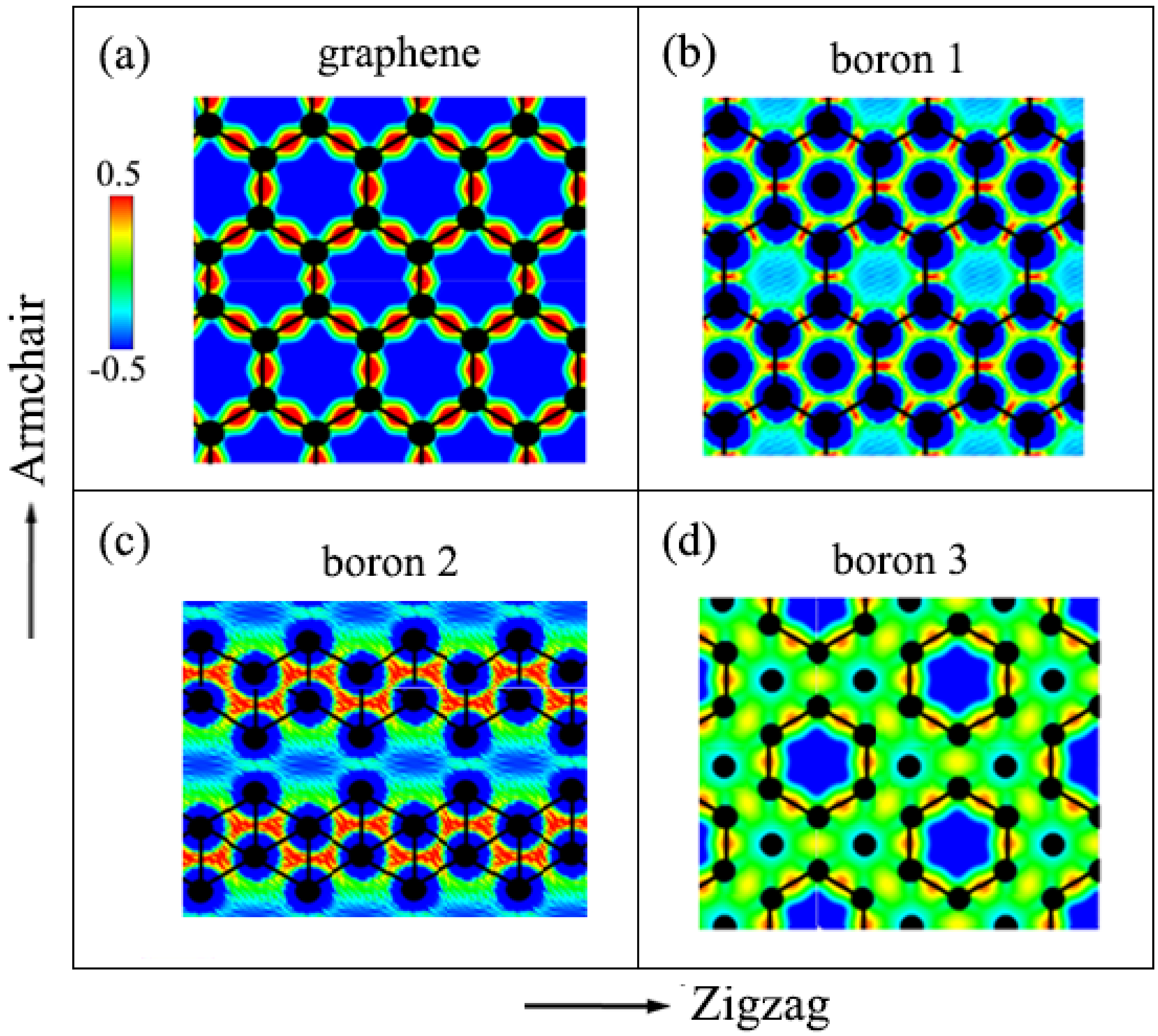

**Figure 5.** BCD contours for (a) graphene, (b) boron 1, (c) boron 2, and (d) boron 3, respectively. Black circles represent the positions of atoms and black lines represent strong bonds.

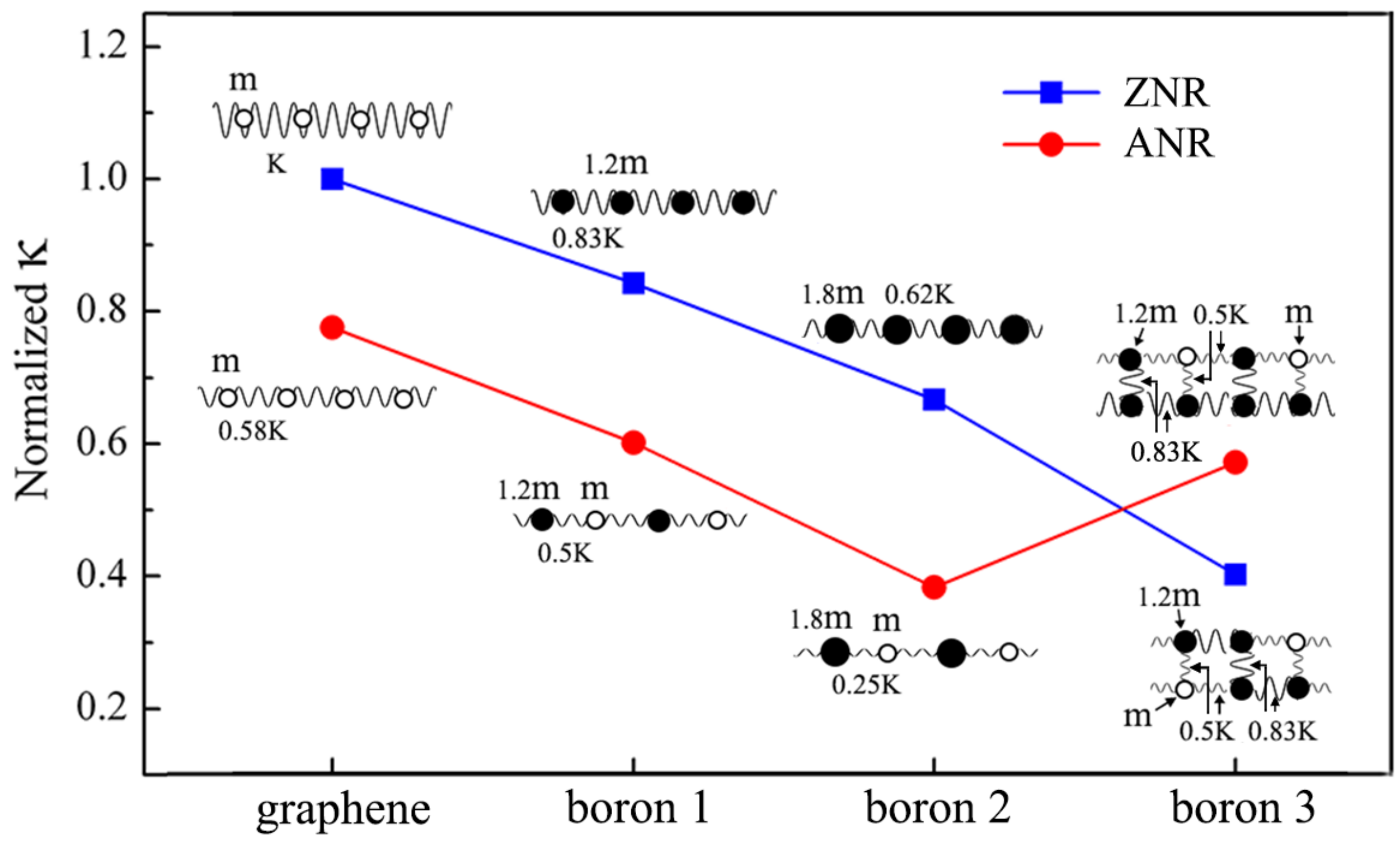

**Figure 6.** Normalized *κ* of simplified models of atomic chains corresponding to graphene and boron nanoribbons. The inset figures are simplified models of atomic chains where the parameters of mass *m* and force constant *K* are shown.

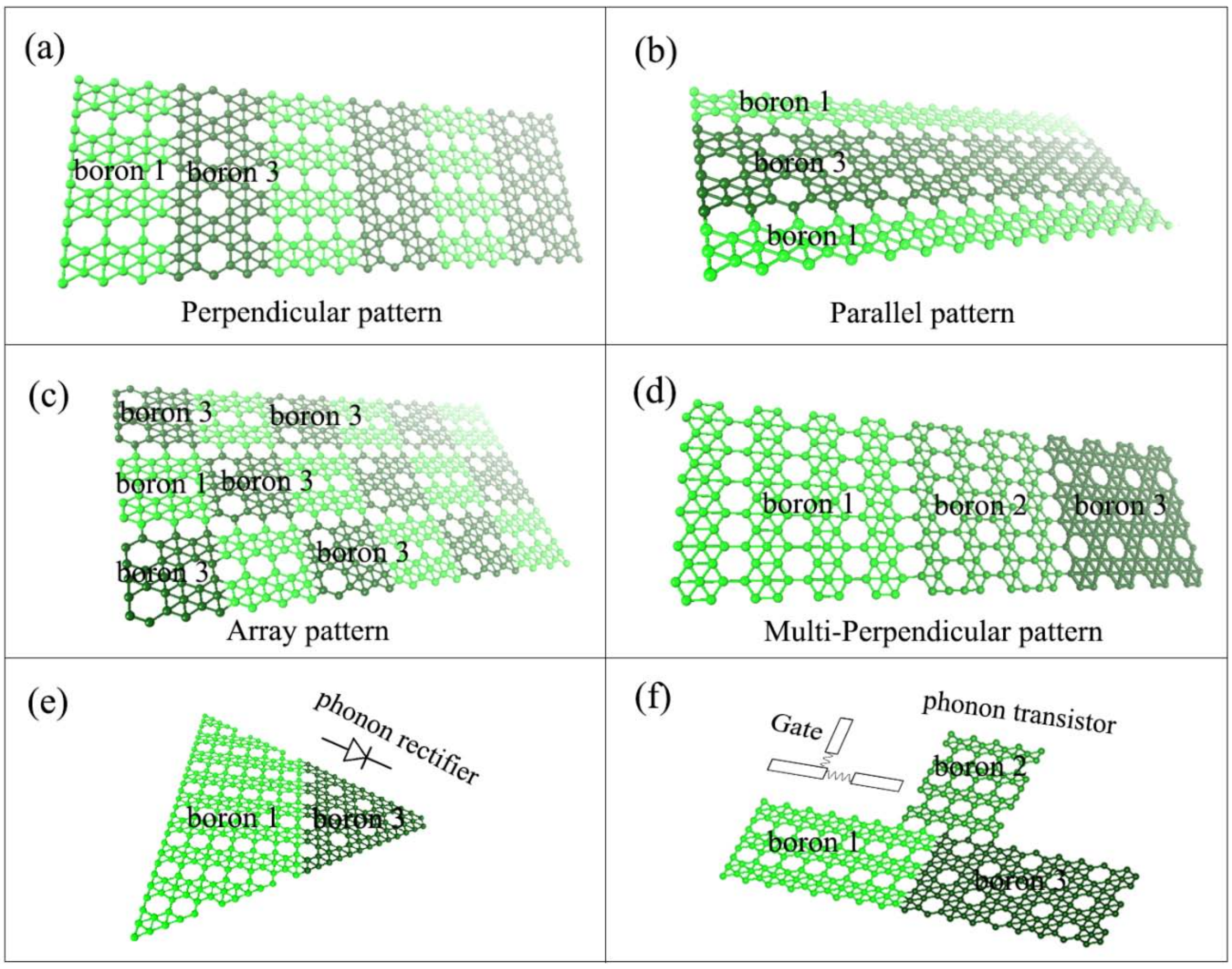

**Figure 7.** Atomic structures of hybrid boron patterns. (a) Perpendicular pattern, (b) parallel pattern, and (c) array pattern consisting of boron 1 and 3. (d) Multi-Perpendicular pattern consisting of boron 1, 2, and 3. (e) A phonon rectifier with triangular shape based on boron 1 and 3. (f) A three-terminal phonon transistor based on boron 1, 2, and 3.

| | $a_1\times10^{-3}$ | $b_1\times10^{-3}$ | $a_2\times10^{-4}$ | $b_2\times10^{-4}$ | $a_3\times10^{-6}$ | $b_3\times10^{-5}$ |
|---|---|---|---|---|---|---|
| B1-ZNR | 2.70 | -2.61 | -1.61 | 8.63 | 1.68 | -2.83 |
| B2-ZNR | 2.25 | -3.14 | -0.85 | 8.70 | -0.63 | -2.86 |
| B3-ZNR | 3.83 | -1.11 | -5.00 | 5.12 | 13.2 | -1.85 |
| B1-ANR | 3.25 | -1.50 | -2.47 | 5.31 | 5.00 | -1.81 |
| B2-ANR | 4.25 | -1.38 | -4.94 | 5.24 | 13.9 | -1.93 |
| B3-ANR | 2.70 | -1.10 | -1.21 | 4.86 | -0.96 | -1.71 |

**Table 1.** The values of fitting parameters $a_i$ and $b_i$ (i =1, 2, 3) in Eq. (6).

# Supplementary Information

# Phonon transport in single-layer Boron nanoribbons

Zhongwei Zhang[1], Yuee Xie[1], Qing Peng[2], and Yuanping Chen[1*]

[1] Department of Physics, Xiangtan University, Xiangtan 411105, Hunan, P.R. China

[2] Department of Mechanical, Aerospace and Nuclear Engineering, Rensselaer Polytechnic Institute, Troy, NY, 12180, USA

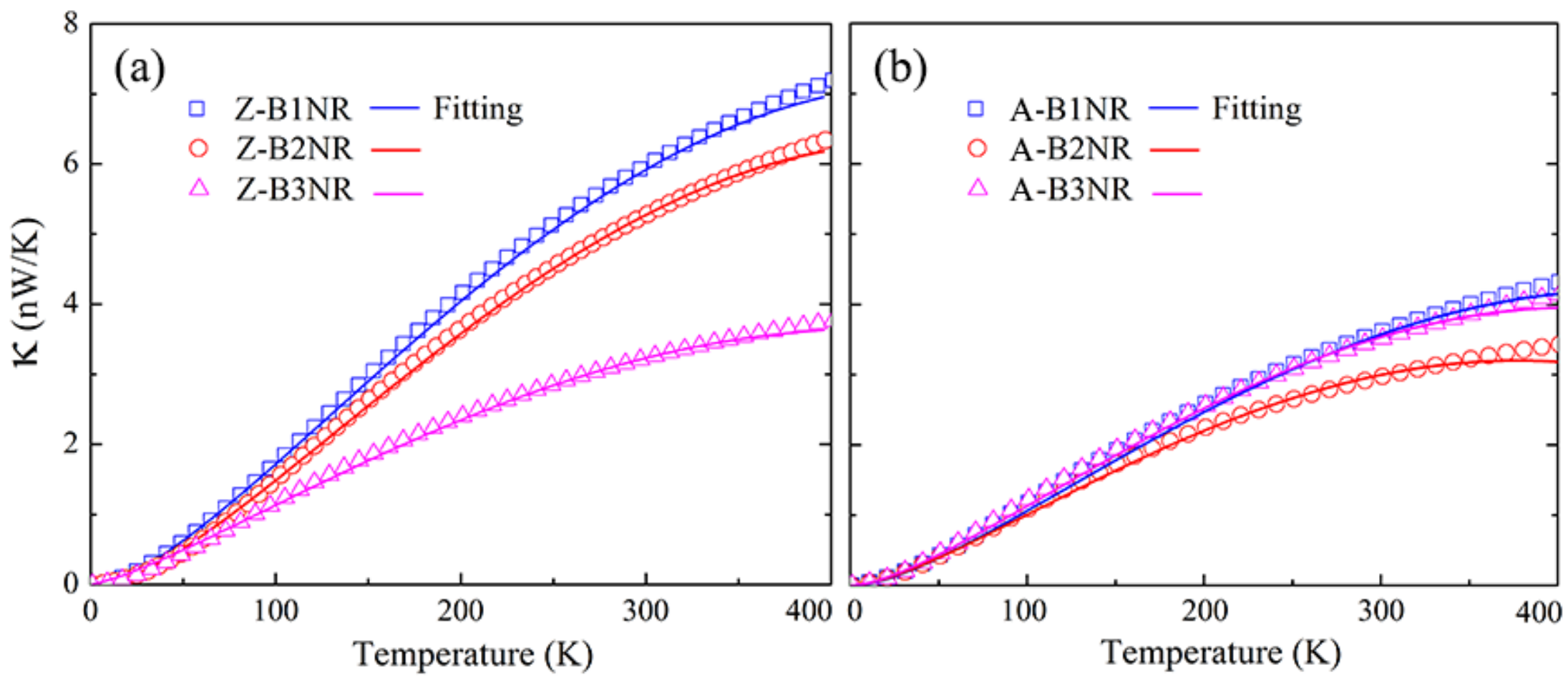

**Figure S1.** Thermal conductance $\kappa$ of (a) B1-ZNR, B2-ZNR and B3-ZNR, and (b) B1-ANR, B2-ANR and B3-ANR as a function of temperature at $W_{Z/A} \approx 5.0$ nm, respectively. The scatters are the calculated data from NEGF, and the solid lines are the fitting curves by Eq. (6) with parameters in Table 1.

As shown in Fig. S1, the thermal conductances of boron nanoribbons with different edges are well fitted by the Eq. (6) with the parameters in Table 1. To analysis the different contributions from $T$, $T^{1.5}$ and $T^2$ terms to phonon transport, in Fig. S2 we have presented the prated $\kappa$ at room temperature and $W_{Z/A} \approx 5.0$ nm. (The used fitting parameters are listed in Table 1.) As one can see, the $T^{1.5}$ term play a great positive contribution to phonon transport in boron nanoribbons of both zigzag and armchair edges, which should dominate the phonon conduct process. Contradictorily, the terms of $T$ and $T^2$ play negative contributions and their values are small. Moreover, the devise anisotropies transport ability can also be reflected in the variation of $T^{1.5}$ term. In boron 1 and 2, the $T^{1.5}$ term's contributions are greater along zigzag edges than the one in armchair edges, correspondingly to the positive thermal conductance anisotropies. While, in boron 3 the contribution has reversed, which lead to a negative anisotropy.

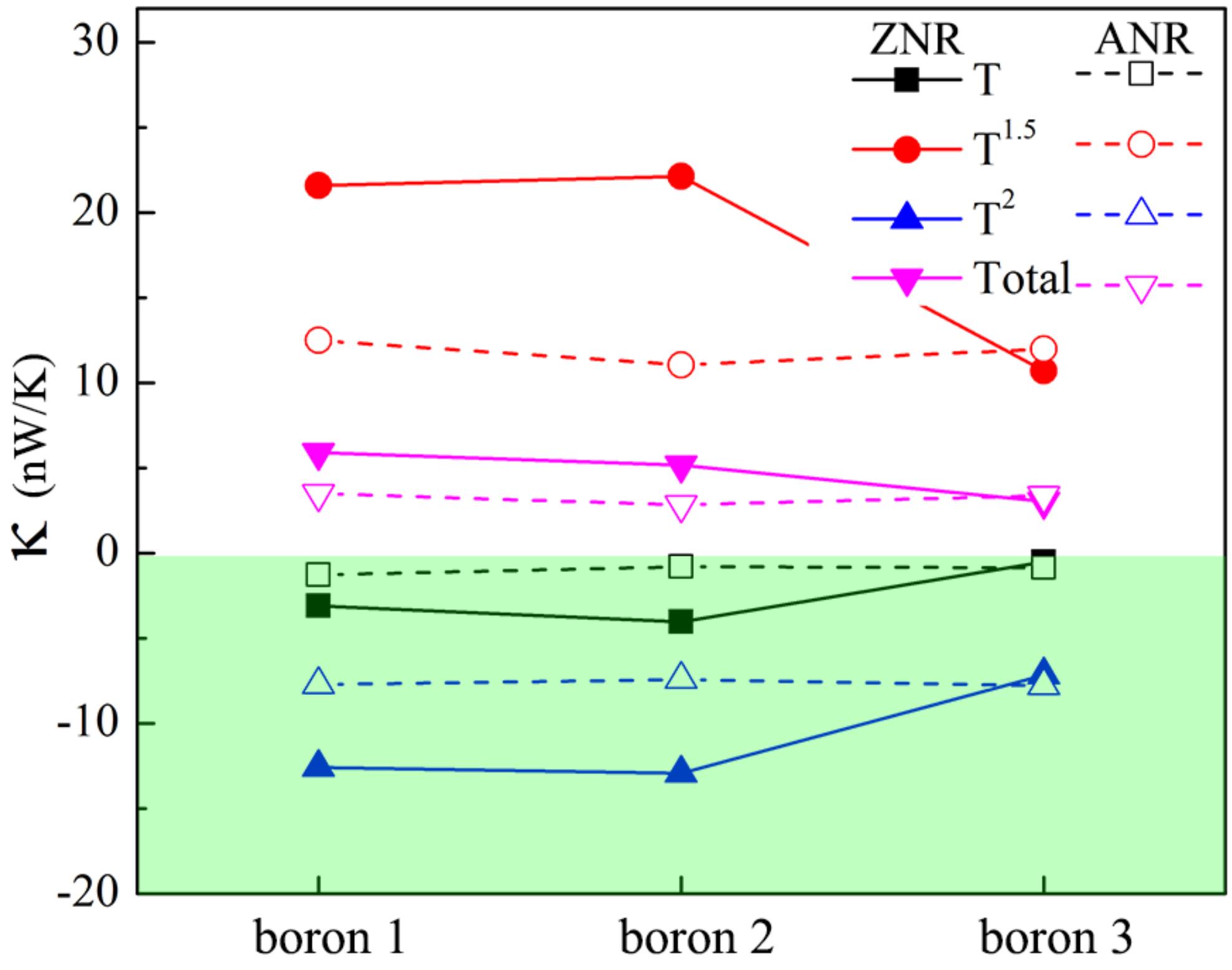

**Figure S2.** Contributions of $T$, $T^{1.5}$ and $T^2$ terms in Eq. (6) of boron 1, 2 and 3 with different edges.